# Modules and Logic Programming[*]

Christophe Fouqueré and Virgile Mogbil

LIPN – UMR7030
CNRS – Université Paris 13
99 av. J-B Clément, F–93430 Villetaneuse
{christophe.fouquere,virgile.mogbil}@lipn.univ-paris13.fr

**Abstract.** We study conditions for a concurrent construction of proof-nets in the framework developed by Andreoli in recent papers. We define specific correctness criteria for that purpose. We first study closed modules (i.e. validity of the execution of a logic program), then extend the criterion to open modules (i.e. validity during the execution) distinguishing criteria for acyclicity and connectability in order to allow incremental verification.

## 1 Introduction

In the last few years, Andreoli [2,3,4] investigated a new style of logic programming aware of resources and aiming at expressing non-determinism, concurrency and (possibly) infinite computations. Logic programming facilitates naturally non-determinism. The other characteristics are out of scope when logic programming is only used for proving goals (possibly by instantiating first order variables). Indeed, proof construction can be goal-directed (and the hypothesis is given as a set of special clauses) or hypothesis-directed (and the goal may be empty). This latter case can be reduced to the first one by contraposing each implication and considering negation of hypotheses as the goal: each resolution is then interpreted as a transformation of the environment. Computation ends when the current environment is empty. As computations build partial proofs, there is no difficulty to take care of (potentially) infinite computations.

It is now a well-known fact that linear logic, a resource-conscious logic, may be used as a programming language after Andreoli's works [1].[1] In the paper just cited, he took a standard approach and his presentation was sequential. More recently, Andreoli takes care of concurrency by switching to proof-nets as this syntax affords a desequentialized presentation of proofs, hence a concurrent way to compute them at the expense of a correctness criterion that guarantees to recover sequentialization, i.e. validity of proofs. In this paper, we search for a generalization of his results in order to have full expressivity. For that purpose, we depart from his approach by adopting a graph point of view. Equivalent to clauses in standard logic programming, *modules*, as graph elements, arise naturally from proof nets. In a few words, associativity, commutativity[2] and focalization lead to polarize formulae, hence to stratify proofnets. Bipolar structures become computational structures as composition of such structures corresponds to some kind of progression rule.

---

[*] Partially supported by ACI NIM project Géométrie du Calcul (GEOCAL), France.

[1] Full first-order linear logic can be used as a programming language. However, we restrict in this paper to propositional multiplicative linear logic.

Andreoli set up this desequentialized framework for middleware infrastructures. In such applications, software agents must satisfy requests or goals by executing concurrently actions on a shared environment. Resources model concrete objects, e.g. documents, or high-level elements, e.g. functionalities. Actions transform the environment by deleting resources and creating sets of results exclusive each other. Andreoli focused on *transitory* proof-structures, i.e. actions always create new resources. Moreover, he imposes prerequisites of actions to be satisfied in order to execute them: the proof construction is exclusively done bottom-up. As we shall see in section 4, these two hypotheses greatly simplify the problem of defining formally conditions under which actions may be undertaken. On the contrary, we constrain neither the structure of modules, nor the application order. It is then possible to define actions that kill resources (e.g. close a branch in a plan, or withdraw a functionality or a resource) or to anticipate consequences of resources still to be acquired. Furthermore, we depart from Andreoli's approach for defining a correctness criterion. His method is based on a computation of domination forests in the spirit of Murawski and Ong's approach [7]. We adopt here a completely different strategy. We define reduction relations in order to get constraints on execution.

The following section gives basic definitions. We formally present modules from elementary ones, graphically and in terms of formulae. We specify in which sense a module is correct, i.e. the computation is allowed. Section 3 is devoted to closed modules. A module is closed when computation ends. Although closed modules are an extreme special case of modules, the methodology we use introduces naturally the way we consider open modules. In a first attempt, we reformulate the resolution rule as a rewriting rule on modules, the Danos-Regnier criterion being used for characterizing correct normal forms. The Danos-Regnier criterion is based on graph properties of proof nets: correct proof structures, i.e. proof nets, are in some sense the connected and acyclic ones. We deduce a correctness criterion for closed modules as our rewriting rule is stable and inverse stable wrt connexity and acyclicity. We define next a modified version of the previous rewriting system that takes care of the parallel structure of modules. Open modules, i.e. modules without constraints, are studied in section 4. We prove that the Danos-Regnier criterion may be extended to open modules seeing that we replace connexity by connectability. We give two rewriting systems as acyclicity and connectability differ fundamentally. These two systems may be viewed as variations over the one we give for closed modules. We end with a study on incrementality wrt composition of modules. In terms of computation, elementary modules compete to modify some current module (the environment). It is then crucial to be able to define rewriting systems that commute with composition.

---

[2] and distributivity when dealing with the additive part of linear logic.

We show that we have to restrict previous rewriting systems for that purpose. However, the rewriting systems have to be splitted into two parts: one commutes with composition, the other is a post-treatment necessary to test correctness of composition.[3]

## 2 Basic definitions

Elementary bipolar modules are our basic blocks. They are interpreted as elementary actions that can take place during an execution. In terms of graph, applying an action is represented as a wire, i.e. composition, of the corresponding (elementary) module onto the current graph. In terms of sequent calculus, this is a resolution step.

**Definition 1 (EBM).** *An* elementary bipolar module *(EBM) $M$ is given by a finite set $\mathcal{H}(M)$ of propositional variables (called hypotheses) $h_i$ and a non empty finite set $\mathcal{C}(M)$ varying over $k$ of finite sets of propositional variables (called conclusions) $c_k^j$. Variables are supposed pairwise distinct.[4] The set of propositional variables appearing in $M$ is noted $v(M)$. Equivalently, one can define it as an oriented graph with labelled pending links and one positive pole under a finite set of negative poles. Its type $t(M)$ and draw are given in the following way:*

$$t(M) = (\bigotimes_i h_i) \multimap (\bigparr_k (\bigotimes_{j_k} c_k^{j_k}))$$

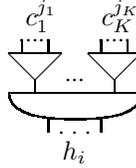

*The set of variables, or equivalently the set of pending links of a module $M$, is called the* border *$b(M)$.*

This specification of modules comes from the fact that connectives are naturally split into two sets: e.g. $\otimes$ is said positive, while $\bigparr$ is negative. Propositional variables are declared positive, and their negation negative. Formulae alternate positive and negative levels up to propositional variables. Moreover, it is possible to flatten proofnets to get bipolar structures related by links on fresh variables as in figure 1. If we notice that a variable and its negation cannot be together linked to negative nodes (it would contradict the correctness criterion), we can always suppose that, say, positive variables are linked to negative nodes. Finally, it may be the case that some bipolar structure (thus beginning with a positive node at bottom) has no negative variable: add then the constant **1**, neutral for $\otimes$. Allowing abusively unary $\otimes$ and $\bigparr$ connectives, these (elementary) bipolar structures are the clauses of our programming language.

---

[3] Complements and some technical proofs are available in the annex of the submission and will be omitted in the final version but remain in a preprint version.

[4] This restriction is taken for simplicity. The framework can be generalized if we consider multisets (of hypotheses and conclusions) instead of sets, and add as required a renaming mechanism: the results in this paper are still true.

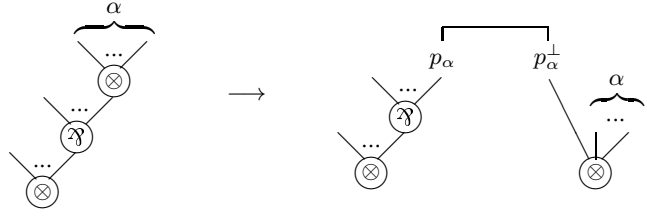

**Fig. 1.** flatten of proof-nets

We thus conveniently suppose that $\bindnasrepma_k F_k = \bigotimes_k F_k = F_1$ when the domain of $k$ is of cardinal 1. Moreover, if the domain of $i$ is empty, $(\bigotimes_i h_i) \multimap C = \mathbf{1} \multimap C$ and if the domain of $j_k$ for some $k$ is empty $(\bigotimes_{j_k} c_k^{j_k}) = \bot$.

*Example 1.* The EBMs $\alpha$ and $\beta$ of respective types $t(\alpha) = a \multimap (b \otimes c)$ and $t(\beta) = b \multimap (d \bindnasrepma (e \otimes f))$ are drawn in the following way:

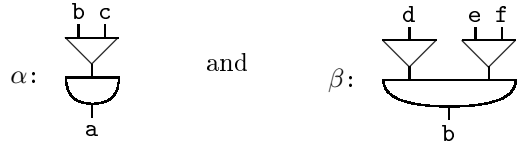

Three kinds of EBMs are of special interest: An EBM is *initial* (resp. *final*) iff its set of hypotheses is empty (resp. its set of conclusions is empty). An EBM is *transitory* iff it is neither initial nor final. Initial EBMs allow to declare available resources, though final EBMs stop part of a computation by withdrawing a whole set of resources. Transitory EBMs are called definite clauses in standard logic programming.

Roughly speaking, a (bipolar) module (BM) is a set of EBMs such that a label appears at most once as a conclusion and at most once as a hypothesis. A label appears as a conclusion and as a hypothesis when two EBMs are linked by this label. As we search for correctness criteria wrt composition of modules (i.e. execution of the program), we give below an inductive definition of bipolar modules.

**Definition 2 (BM).** *A bipolar module (BM) $M$ is defined with hypotheses $\mathcal{H}(M)$, conclusions $\mathcal{C}(M)$, and type $t(M)$, inductively in the following way:*

- *An EBM is a BM.*
- *Let $M$ be a BM, and $N$ be an EBM, let $I = \mathcal{C}(M) \cap \mathcal{H}(N)$, their composition wrt the interface $I$, $M \circ_I N$ is a BM with :*
  - *the multiset of hypotheses $\mathcal{H}(M) \cup (\mathcal{H}(N) - I)$*
  - *the multiset of conclusions $(\mathcal{C}(M) - I) \cup \mathcal{C}(N)$*
  - *the type $t(M) \otimes t(N)$*
  - *the variables $v(M) \cup v(N)$*

The informal explanation given before is more general than this definition because we define BM incrementally. However, we abusively do not consider these differences in the following as properties will be proven in the general case. The interface will be omitted when it is clear from the context. Note that the interface may be empty: it only means that two computations are undertaken, currently without any shared resources. A BM may not correspond to a valid computation: e.g. we do not want to accept that some action uses two resources in disjunctive situation! Correctness has obviously to be defined wrt the underlying Linear Logic as we do below. Finally, note that when a BM is correct, it represents the history of the computation whereas its conclusion is the current available environment.

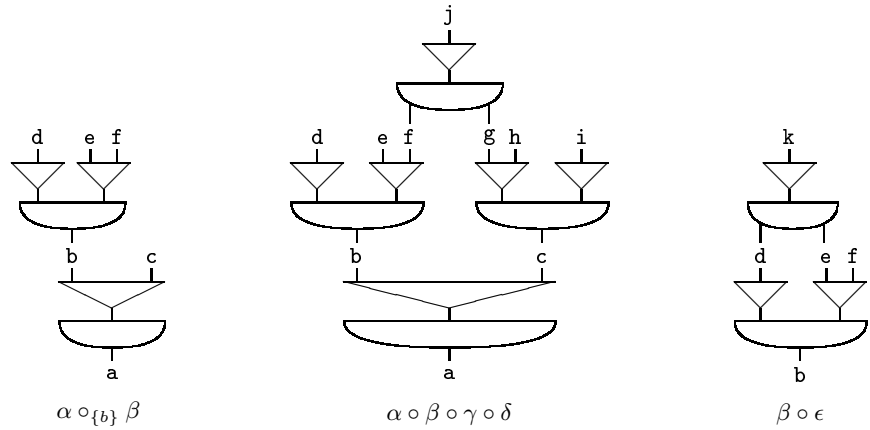

**Fig. 2.** examples 2 and 3

*Example 2.* The composition of the EBMs $\alpha$ and $\beta$ is the BM $\alpha \circ_{\{b\}} \beta$ drawn in figure 2. Its type is $t(\alpha) \otimes t(\beta)$.

**Definition 3 (Correctness (wrt sequentialization)).** *Let $M$ be a BM, $M$ is correct iff there exists a formula $C$ built with the connectives $\otimes$ and $\mathbin{\bindnasrepma}$, and the variables $\mathcal{C}(M)$ such that the sequent $\mathcal{H}(M), t(M) \vdash C$ is provable in Linear Logic.*

*Example 3.* Let us give two more BMs $\delta$ and $\gamma$ of respective types $(f \otimes g) \multimap j$ and $c \multimap ((g \otimes h) \mathbin{\bindnasrepma} i)$.

- The following sequent is provable in LL: $a, t(\alpha \circ \beta \circ \gamma \circ \delta) \vdash d \mathbin{\bindnasrepma} (e \otimes j \otimes h) \mathbin{\bindnasrepma} i$. The (correct) BM $\alpha \circ \beta \circ \gamma \circ \delta$ is drawn in the figure 2.
- Let $\epsilon$ be an EBM of type $(d \otimes e) \multimap k$, the BM $\beta \circ \epsilon$ is not correct. Note the trip through $d$ and $e$ in the figure 2.

As we shall focus first on characterizing correctness on closed modules, and then generalize our results to open modules, we adjoin to the term *correct* the kind of modules we speak of, e.g. c-correct when the module is closed, o-correct when it is open.

## 3 Closed modules

A closed module is a BM where the sets of hypotheses and conclusions are empty. Correctness of closed modules may be tested either in sequent calculus or by means of proofnets. We use this latest representation in this section. Girard in his seminal paper [6] gave a parallel syntax for multiplicative linear logic as oriented graphs called *proof-nets*. A *correctness criterion* enables one to distinguish sequentializable proof-structures (say such oriented graphs) from "bad" structures. The reader may find in [5] the definitions of proof structures and switchings. One generalizes this definition to n-ary connectives in the obvious way (taking care of associativity and commutativity of $\otimes$ and $\invamp$) in place of standard binary ones. One modifies in the same way the definitions of switching introducing generalized switches. In particular a $n$-ary $\invamp$ connective has $n$ switched positions. One still can define switched proof-structures and a criterion generalizing Danos-Regnier correctness criterion: A closed module $M$ is DR-correct iff for all generalized switches $s$ on $M^*$, $s(M^*)$ is acyclic and connected, where $M^*$ is the proof structure associated to $t(M)^\perp$.[5] We immediately have the following proposition as a corollary of the DR-criterion theorem:

**Proposition 1 (c-correction).** *Let $M$ be a closed module,*

$M$ *is c-correct iff $t(M) \vdash$ is provable in Linear Logic,*
         *iff $M$ is DR-correct.*

Remember that the equivalent (binary) Danos correctness criterion may be implemented by means of a contraction relation on proof structures. However, intermediate reduced structures may not be describable in terms of (bipolar) modules. Moreover such a contraction relation does not take advantage of the incremental definition of modules as a composition of elementary bipolar modules. A first idea consists in representing the resolution step (implicit in EBMs composition) in terms of modules. We first give below such a (small step) reduction rule that is stable wrt correctness with $\underset{\cup}{\mathsf{Y}}$ as the correct normal form, where $\underset{\cup}{\mathsf{Y}}$ denotes the terminal EBM (i.e. smallest final and initial). We give then a second proposal that takes care of the focalization property. Though a resolution step reduces one variable, this second formulation uses as a whole the structure of a module thanks to focalization.

Let $\leadsto_\Theta$ be the transitive closure of the following relation defined on literals of a proof-structure $\Theta$: let $u$ and $v$ be two literals of $\Theta$, $u \leadsto_\Theta v$ iff $u^\perp$ and $v$ are in the same subtree with root $\otimes$ of the formula corresponding to $\Theta$. We note $u \rightsquigarrow v$ when there is no ambiguity. In the following, we consider proof-structures modulo neutrality of the constant 1 and associativity of connective ⅋.

---

[5] We abusively note $s(M)$ in place of $s(M^*)$ in the following.

**Definition 4 (Small step reduction rule).** *Let $\rightarrow$ be the reduction relation given by: if $\forall v$ a literal of $\psi$, $v \not\rightsquigarrow x^\perp$ then*

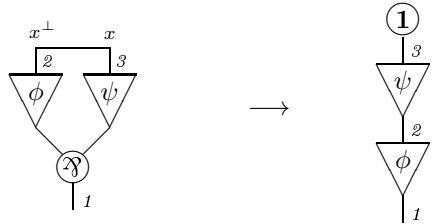

**Theorem 1 ((small steps) Correctness criterion).** *Let $M$ be a closed BM, $M$ is correct iff $M^* \rightarrow^* 1$.*

Briefly speaking, one can prove that the relation $\longrightarrow$ and the inverse relation are stable wrt DR-correctness by induction over the height of $\psi$. One may want to get rid of the (global) condition in favor of a local condition. This is possible thanks to the structure of modules. Suppose $M$ is a correct closed module, then one may define an equivalent proof-net by sufficiently adding fresh variables as described in the introduction. It is easy to prove that the constraint is satisfied by $x$ or $x^\perp$ for each variable $x$. However, the reduction system being not strongly confluent, a reduction on a variable may lead to a proof structure on which the condition is not always satisfied. There are two cases where this does not happen: either all variables on a tensor have their negation on the same ⅋, or the converse interchanging ⅋ and $\otimes$. The following (big step) reduction relation $\twoheadrightarrow$ with two rewrite rules uses this fact. Note that this system is confluent and terminates.

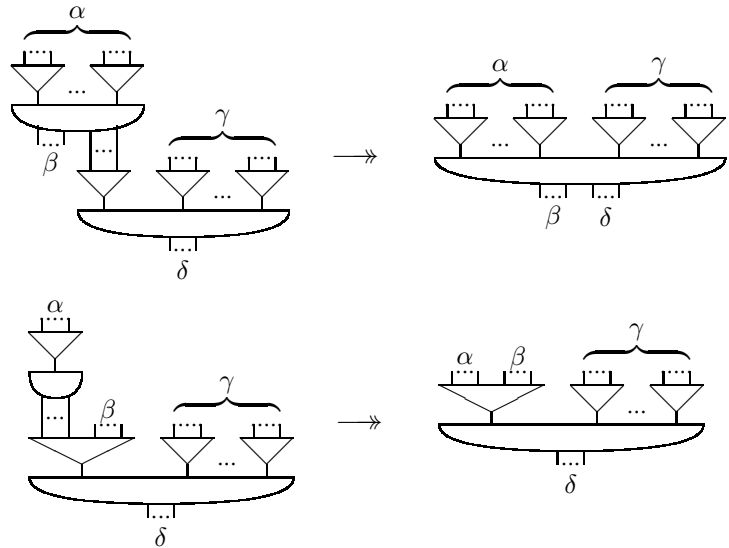

**Proposition 2 (Stability).** *Let $M$ and $N$ be two closed modules and $M \twoheadrightarrow N$, $M$ is c-correct iff $N$ is c-correct.*

*Proof.* One can define a function from left switched module onto right switched module stable wrt acyclicity, connexity, and the inverse properties. □

**Theorem 2.** *A closed module $M$ is c-correct iff $M \twoheadrightarrow^* \underset{\cup}{\mathsf{Y}}$.*

*Proof.* As the reduction rules are stable wrt correctness, it remains to prove that a correct non-terminal closed module $M$ can always be reduced. We define a partial relation on negative poles: a negative pole is smaller than another one if there exists a positive pole st the first negative pole is linked to the bottom of the positive pole and the second negative pole is linked to the top of the positive pole. We consider the transitive closure of this relation.

If maximal negative poles do not exist then there exists at least one cycle in the module alternating positive and negative poles. We can then define a switching function on the module (choosing the correct links for negative poles) st the switched module has a cycle. Hence contradiction.

So let us consider one of the maximal negative pole, and the corresponding positive pole. We remark that such a negative pole has no outcoming links (the module is closed and the negative pole is maximal). If the positive pole has other negative poles, we can omit the maximal negative pole by neutrality. Otherwise, let us study the incoming negative poles.

If there is no such incoming link, then $M$ is the terminal module. If each incoming negative pole has at least one link going to another positive pole, then one can define a switching function using for each of these negative poles one of the link that does not go to the positive pole we considered first. Hence the switched module is not connected (there are no outgoing links). Hence contradiction. So there exists at least one incoming negative pole with the whole set of links associated to the positive pole: the first rule applies and we are finished. □

## 4 Open modules

### 4.1 O-correction

We focus in this section on open modules. An *open module* is a possibly non closed BM. The bigstep reduction relation presented in the previous section is not sufficient to characterize again correction of open module. Let $U$ be the first module of the next example. One cannot apply on $U$ a bigstep reduction on the negative pole with variable $a$ as this pole remains in the normal form though $U$ seems correct. The c-correctness theorem 2 is no more valid for open modules.

Correctness of open modules is defined wrt correctness of closed extensions. A *closure* $\overline{M}$ of an open module $M$ is a closed module such that $M$ is a submodule of $\overline{M}$. As a BM is a graph with pending edges, one defines submodules and induced modules as expected. We use the notation $\widetilde{M}$ for the module $\overline{M}$ without $M$ but with border $b(M)$. In the following example $\overline{U}$ is a closure of $U$ st $\widetilde{U}$ is the right module. The composition of $U$ with a set of only initial/final EBMs is a closure too.

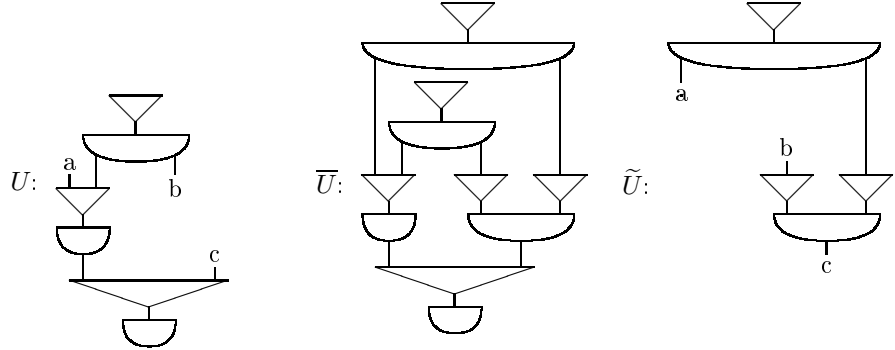

An open module $M$ is *o-correct* iff there exists a c-correct closure of $M$. The open module $U$ of the example is o-correct because the given closure is c-correct. Note that there is no other c-correct closure. Hence it is not possible in general to split the problem of finding a closure into finding a completion by initial modules and final modules. In the previous section, we defined a rewriting system able to test the correctness of a closed module. As this system is stable wrt connexity and acyclicity, it is invariant wrt the Danos-Regnier criterion. In order to take care of open modules, we extend connexity to connectability (acyclicity is treated easily) and prove that connectability and acyclicity are necessary and sufficient for o-correctness. However, we are not able to define a single rewriting system that commutes with composition. An open module $M$ is *acyclic* if for all generalized switches $s$ on $M$, $s(M)$ is acyclic. Note that a submodule of an acyclic module is obviously acyclic.

An open module $M$ is *connectable* iff there exists a connected closure $\overline{M}$ st $\widetilde{M}$ is acyclic. As a connected closed module is already connectable (just take itself as closure), the connectability is an extension of the connexity property. We give an equivalent definition: an open module $M$ is connectable iff the closed module $M \circ F$ is connected where $F$ is a *full connector* EBM for $M$, i.e. $F$ has as hypotheses the set of conclusions of $M$, is final if $M$ has no hypothesis or has a negative pole with one conclusion for each of its hypotheses. In fact if there exists a connected closure $\overline{M}$ then $M \circ \widetilde{M}$ is connected. So *a fortiori*, $M \circ F$ is connected. The converse comes from the definition.

**Theorem 3 (o-correction).** *An open module $M$ is o-correct iff $M$ is acyclic and connectable.*

*Proof.* By definition o-correction implies acyclicity and connectability.
If $M$ is acyclic and there exists a connected closure $\overline{M}$ st $\widetilde{M}$ is acyclic then by induction on the number of cycles of $\overline{M}$, one can construct an acyclic and connected closure of $M$.
If there is a cycle $\sigma$ in $\overline{M}$ then by hypothesis $\sigma \cap b(M) \neq \emptyset$. Suppose there exists a hypothesis of $M$ $h \in \sigma \cap b(M)$, one defines $N$ to be $\widetilde{M}$ where we substitute a fresh label $h'$ to $h$. Let $N'$ be the composition of the initial EBM of border $\{h\}$, the final EBM of border $\{h'\}$ and $N$. $M \circ N'$ has one cycle less than $\overline{M}$ and is a connected closure.

Otherwise the elements of $\sigma \cap b(M)$ are conclusions of $M$. Let $c$ be such a conclusion. We consider the following cases:

- if $c$ in $\sigma \cap b(M)$ is the only conclusion of a negative pole $n$, then one can do the same thing as in the previous case.
- else let $d$ be a conclusion in $\sigma \cap b(M)$ distinct from $c$ of $n$. One renames $c$ (resp. $d$) in $\widetilde{M}$ in $c'$ (resp. $d'$) to get $N$. One defines also an EBM $D$ with one conclusion $d'$ and two hypotheses $c$ and $d$, and an initial EBM $E$ with conclusion $c'$. Then $X = M \circ D \circ E \circ N$ is a connected closure of $M$ and $D \circ E \circ N$ is acyclic. Hence $X$ is a connected closure of $M \circ D$ and $E \circ N$ is acyclic. We suppressed the cycle $\sigma$. However, it may be the case that there were a cycle through $d$ and $D$ doubles it ! For that purpose, we transform $M$ to get rid of this extra cycle. Let $M'$ be $M$ where we identify the two edges labelled $c$ and $d$ in one labelled $d'$. Then $M' \circ E \circ N$ is a connected closure of $M'$ and $E \circ N$ is acyclic. Moreover the number of cycles in $M' \circ E \circ N$ is one less than in $\overline{M}$. Thus there exists $N'$ acyclic such that $M' \circ N'$ is c-correct. Hence $M \circ D \circ N'$ is c-correct. □

### 4.2 Acyclicity and Connectability Criteria

**Acyclicity.** An open module $M$ *restricted to the subset $I$* of $b(M)$ is the subgraph of $M$ where we omit pending edges not in $I$. We denote it $M\!\!\upharpoonright_I$. Unformally an open module $M$ restricted to $I$ is a submodule of border $I$. The restriction of an open module to the empty set is a closed module. Restriction gives naturally an equivalent definition of acyclicity for open modules: an open module $M$ is *acyclic* iff the closed module $M\!\!\upharpoonright_\emptyset$ is acyclic. Hence the proposition given in the previous section applies:

**Proposition 3 (acyclicity).** *An open module $M$ is acyclic if $M\!\!\upharpoonright_\emptyset \twoheadrightarrow^* \underset{\cup}{\mathsf{Y}}$.*

*Proof.* $M\!\!\upharpoonright_\emptyset$ is a closed module and $M\!\!\upharpoonright_\emptyset \twoheadrightarrow^* \underset{\cup}{\mathsf{Y}}$ then by (inverse) stability of acyclicity $M\!\!\upharpoonright_\emptyset$ is acyclic. $M$ is then acyclic. □

Note that the converse is not true, otherwise acyclic closed modules would be correct! A way to characterize acyclicity by means of a reduction relation is to enlarge the reduction $\twoheadrightarrow$ (quotienting the set of normal forms). Splitting the negative poles suffices to continue reduction until we get a non-empty set of $\underset{\cup}{\mathsf{Y}}$: closing modules may link disjoint connected components. It is then obvious to deduce a necessary and sufficient condition for acyclicity.

Andreoli considered in [4] only transitory proof-structures. A *transitory proof-structure* is equivalent to a BM without hypothesis[6] such that negative poles have always conclusions and obtained by a bottom-up composition of EBMs. As negative poles have pending edges, there is always a way to connect it to other parts of the module: if a transitory module $M$ is acyclic then $M$ is connectable. Hence a transitory module $M$ is o-correct iff $M$ is acyclic. The reduction relation we give to test acyclicity can be considered as an alternative to Andreoli's method.

---
[6] In fact, there may be hypotheses in built modules but these are unused.

**Connectability : a contraction relation.** The proof of the correctness of the big step reduction relation for closed modules gives the keys for finding a connectability property that relies on the structure of an open module (and not on the modules candidate to close it !). Proof of theorem 2 is based on reducing first maximal negative poles. In the case of open modules, maximal elements may have pending edges that should be connected in the closure. But we notice that we keep connectability if we replace the whole set of pending edges for such an element by just one pending edge. With this in mind, we consider the following (non oriented) contraction relation on (contracted) modules:

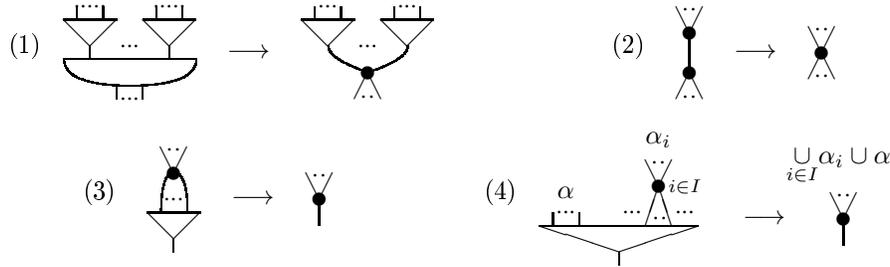

Rule (4) is restricted to cases where the negative pole is such that for all $i \in I$, $\alpha_i \cap b(M) \neq \emptyset$ and $\alpha \subseteq b(M)$ where $b(M)$ is the set of pending edges of $M$, i.e. the border set. The sets $I$ and $\alpha$ may be empty. We denote by $\rightarrow_c^1$ one rewriting step and by $\rightarrow_c$ the reflexive and transitive closure of $\rightarrow_c^1$. We call *contracted node* a black node. Note that rule (4) is simply the rewriting of a negative pole in a contracted node if the condition is satisfied. Thus acyclicity is not preserved but connectability is.

**Proposition 4.** *The relation $\rightarrow_c$ is strongly confluent and terminates.*

*Proof.* The first rule acts just as a mark. We can forget it: it is just for convenience. Each rule applies locally and strictly decreases the number of negative poles and contracted nodes. The rules are disjoint except for a pair of negative poles linked by the same contracted node $i_0$ for which rule (4) can be applied (it is a trivial case), and except for a left member of rule (3) st rule (4) applies too: in this case results are identical. □

We extend the notions of switching to modules with contracted nodes: contracted nodes are treated as positive poles. Acyclicity, connexity, closure and connectability are extended in the same way. As in section devoted to closed modules, our strategy consists in characterizing amongst normal forms of this relation the correct ones, and prove stability of, say, connectability. Let $M$ be an open module and $f$ the corresponding normal form. By definition if $f$ does not contain a negative pole then $f$ is a set of contracted nodes $\{n_j\}_{j \in J}$ st all pending egdes are in $b(M)$. We use the notation **cc** for a set of contracted nodes $\{n_j\}_{j \in J}$ st for all $j \in J$ $n_j$ has at least one edge in the border $b(M)$ except if $|J| = 1$. If $f$ contains a negative pole $N$ then, $f$ being a normal form of relation $\to_c$, rule (4) does not apply on $N$. Hence the set $I$ as defined by rule (4) is st there exists $i_0 \in I$, $\alpha_{i_0} \cap b(M) = \emptyset$. Moreover this contracted node $i_0$ is linked to hypotheses of negative poles $\{h_l\}_{l \in L}$ and to conclusions of only negative poles $\{c_k\}_{k \in K}$ st each of them has other conclusions $\beta_k \neq \emptyset$ not linked to $i_0$ (otherwise rule (2) applies for such nodes):

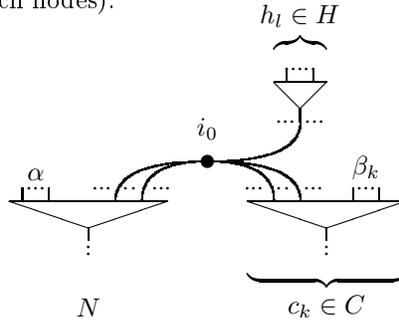

If we suppose the negative pole $N$ is a maximal one (i.e. $H = \emptyset$), there is a switching (on $\alpha$ or on some $i \neq i_0$ and on one of each $\beta_k$) st $f$ (as closures of $f$) is not connected. Thus $f$ is not connectable.

*Example 4.* The following subforms imply not connectability:

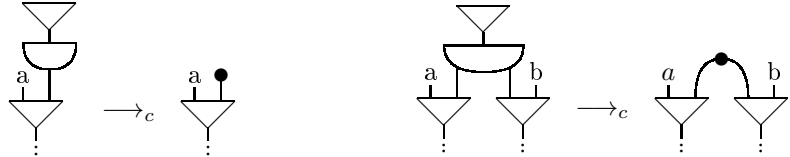

**Proposition 5 (stability).** *Connectability is stable wrt (resp. inverse) contraction rules.*

*Proof.* The three first rules satisfy obviously stability and inverse stability. Let $M$ be an open module st $M \to_c^1 M'$ by the contraction rule (4) and there exists $\overline{M}$ connected and $\widetilde{M}$ acyclic. Obviously $M' \circ \widetilde{M}$ is connected. Concerning inverse stability, let $M$ be an open module st $M \to_c^1 M'$ by the contraction rule (4) and let $F$ be a full connector EBM for $M'$. Note that $b(M') = b(M)$. The connectability of $M'$ implies that $M' \circ F$ is connected. Wrt rule (4), because for all $i \in I$, $\alpha_i \cap b(M) \neq \emptyset$ and $\alpha \subseteq b(M)$, for every switches $s$, $s(M \circ F)$ is connected too. □

By stability and inverse stability of connectability we have:

**Theorem 4.** *Let $M$ be an open module, $M$ is connectable iff $M \to_c$ **cc**. Hence an open module $M$ is o-correct iff $M$ is acyclic and $M \to_c$ **cc**.*

## 5 Composition of modules

In the sequel we discuss an incremental criterion to test the composition of an open module with an EBM. Let $M$ be an o-correct open module and $E$ an EBM st $b(M) \cap b(E) \neq \emptyset$ (otherwise the test is easy). As seen above, acyclicity and connectability, hence o-correctness, of $M$ may be decided by computing normal forms. Our aim is to decide the o-correction of the composition $M \circ E$ 'incrementally' i.e. not directly but o-correctness of $M$ being given. From the previous section we have:

$$M \text{ is o-correct iff } M\!\downarrow_\emptyset \twoheadrightarrow^* \mathsf{Y} \text{ and } M \to_c \mathbf{cc}$$

Because of the restriction of $M$ to the empty border, the acyclicity condition given above does not commute with composition. It is the same for connectability: even if there is preservation of the border with $\to_c$, a choice is made for the completion of $M$ which may be different from the way composition with $E$ is done. For example we have:

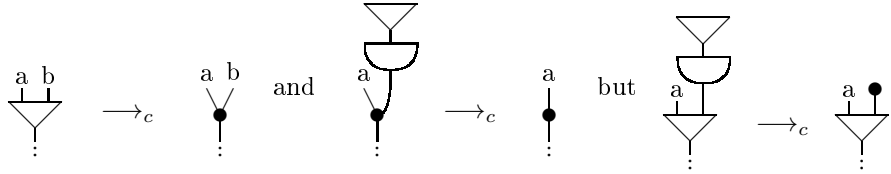

In the sequel we show that if we release the restriction operation we can incrementally manage acyclicity. The relax of the (implicit) completion in the rewriting rules dealing with connectability gives also an incremental (but not so convenient) criterion for connectability.

### 5.1 Incremental acyclicity

Note that the restriction to empty set is stable wrt the reduction $\twoheadrightarrow$ i.e. if $M$ is an open module st $M \twoheadrightarrow^1 N$ then $M\!\downarrow_\emptyset \twoheadrightarrow^1 N\!\downarrow_\emptyset$. Hence an incremental test for acyclicity follows:

**Proposition 6.** *Let $M$ be an open module st $M \twoheadrightarrow^* f$ and $E$ an EBM. $M \circ E$ is acyclic if $(f \circ E)\!\downarrow_\emptyset \twoheadrightarrow^* \mathsf{Y}$.*

*Proof.* If $M \twoheadrightarrow^* f$ then $(M \circ E) \twoheadrightarrow^* (f \circ E)$. Following previous remark, $(M \circ E)\!\downarrow_\emptyset \twoheadrightarrow^* (f \circ E)\!\downarrow_\emptyset$. Thus if $(f \circ E)\!\downarrow_\emptyset \twoheadrightarrow^* \mathsf{Y}$ then $(M \circ E)\!\downarrow_\emptyset \twoheadrightarrow^* \mathsf{Y}$. □

### 5.2 Contraction relation (without completion)

We consider the rewriting system given to test connectability where rule (4) is restricted to the following degenerated case ($\alpha = I = \emptyset$ and application of rule (2)):

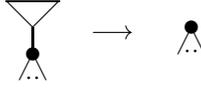

We denote by $\to_w^1$ one rewriting step and by $\to_w$ the reflexive and transitive closure of $\to_w^1$. As it is a subsystem of the previous one, the relation $\to_w$ terminates and is still strongly confluent (there is only trivial independant pairs).

We study the normal forms. By definition an open module contracts in a normal form composed with only contracted nodes or contracted modules where each negative pole $N$ is of the following form:

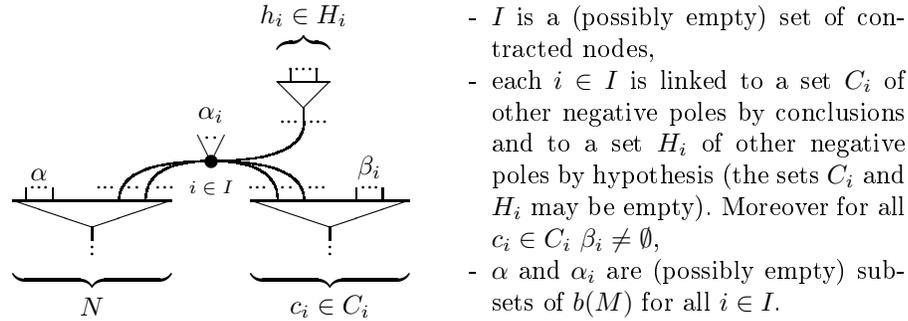

- $I$ is a (possibly empty) set of contracted nodes,
- each $i \in I$ is linked to a set $C_i$ of other negative poles by conclusions and to a set $H_i$ of other negative poles by hypothesis (the sets $C_i$ and $H_i$ may be empty). Moreover for all $c_i \in C_i$ $\beta_i \neq \emptyset$,
- $\alpha$ and $\alpha_i$ are (possibly empty) subsets of $b(M)$ for all $i \in I$.

We focus on the two possible forms of negative pole:

- there exists $i_0 \in I$ st $\alpha_{i_0} = H_{i_0} = \emptyset$. We denote such forms by **notcc**.
- for all $i \in I$, $\alpha_i \neq \emptyset$ or $H_i \neq \emptyset$. These negative poles may be considered in the previous system $\to_c$.

If a normal form has no negative poles then it is a set of contracted nodes. We add to the **notcc** forms the case where there is at least one contracted node without pending edges and other nodes.

In order to compare these normal forms with the normal forms of $\to_c$ observe that: (i) by definition of normal forms, if $I = \emptyset$ then $\alpha \neq \emptyset$, and if $I \neq \emptyset$ then $\mid I \mid \geqslant 2$ or $\alpha \neq \emptyset$, (ii) for all $i \in I$ for all $c_i \in C_i$ we have $\beta_i \neq \emptyset$. It follows that if a normal form $g$ wrt $\to_w$ of an open module $M$ contains a **notcc** subform then there is a generalized switch st $g$ is not connected. The stability of connexity wrt $\to_w$ being given, $M$ is not connected (neither its closures), thus not connectable.

Remark that the **notcc** forms are already in the previous system: they are normal forms which are not the **cc** forms! In fact the **notcc** subforms are invariant wrt the previous system $\to_c$. Moreover as inverse stability of connectability is easily proven, we have:

**Theorem 5.** *Let $M$ be an open module, $M$ is connectable iff $M \to_w g$ st $notcc \notin g$.*

*Proof.* Let $M$ be st $M \to_w g$. If $\mathbf{notcc} \in g$ then $g$ is not connected (neither its closures) and by stability of connexity $M$ is not connectable. Conversely, if $\mathbf{notcc} \notin g$ then $g \to_c \mathbf{cc}$ by invariance of $\mathbf{notcc}$ wrt $\to_c$. By theorem 4, $g$ is connectable. The result is obtained by inverse stability of connectability wrt $\rightarrow_w$.
□

Hence, an open module $M$ is o-correct iff $M$ is acyclic and $M \rightarrow_w g$ st **notcc** $\notin g$. By confluence property and theorem 5 we have an incremental test: Let $M$ be a connectable open module st $M \rightarrow_w g$ and $E$ an EBM st $b(M) \cap b(E) \neq \emptyset$. We have:

$$M \circ E \text{ is connectable iff } f \circ E \rightarrow_w g \text{ st } \textbf{notcc} \notin g.$$

### 5.3 A test for composition

Testing the composition of an EBM $E$ on a correct module $M$ may be done in the following way. We associate to such a module $M$ a pair $(f, g)$ such that $M \twoheadrightarrow^* f$ and $M \rightarrow_w g$. We compute the pair $(f', g')$ associated to $M \circ E$: $f \circ E \twoheadrightarrow^* f'$ and $g \circ E \rightarrow_w g'$. Then $E$ may be plugged onto $M$, i.e. the composition is correct, iff $f'|_\emptyset \twoheadrightarrow^* \underset{\cup}{\curlyvee}$ and **notcc** $\notin g'$. This test may be implemented in such a way that pre-computations are done in $M$ in order to optimize the test. Moreover this pre-computation allows for a concurrent treatment for testing composition by only locking a reduced part of the module $M$.

## 6 Conclusion

Concurrent construction of proof-nets allows for a new approach in designing concurrent logic programming languages. In the framework developed by Andreoli in recent papers, we first presented a criterion for testing the correctness of closed modules (i.e. validity of the execution of a logic program), then we extended the criterion to open modules after proving that correctness of open modules reduces to testing acyclicity and connectability. Furthermore, criteria for acyclicity and connectability lead naturally to incremental verification.

# 7 Annex

## 7.1 Proofnets

Girard in his seminal paper [6] gave a parallel syntax for multiplicative linear logic as oriented graphs called *proof-nets*. A *correctness criterion* enables one to distinguish sequentializable proof-structures (say such oriented graphs) from "bad" structures.

**Definition 5 (Proof structure).** *A* proof structure *is a graph whose vertices are labelled with formulae and built from the following links (i.e. graphs):*

- *Axiom-link (two conclusions, no premise)*
- *Cut-link (two premises, no conclusion)*
- *$\otimes$-link (two premises, one conclusion)*
- *$\invamp$-link (two premises, one conclusion)*
- *$1$-link (no premise, one conclusion)*

*and every occurrence of a formula is a premise of at most one link and is a conclusion of exactly one link.*

For every link $l$, a set $\mathcal{S}(l)$ of graphs called Switching positions *is given*. $\mathcal{S}(l) = \{l\}$ *except when $l$ is a $\invamp$-link*. $\mathcal{S}(\invamp\text{-link})$ *is defined by the two following switches:*

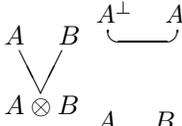

*A* switching *$s$ of a proof-structure $\Theta$ is a function which associates a switching position $s(l) \in \mathcal{S}(l)$ to every link $l$ of $\Theta$. The* switched proof-structure *$s(\Theta)$ is the graph with vertices the formulae labelling $\Theta$, and edges the ones given by the switching function $s$.*

*A* Danos-Regnier proof-net *is a proof-structure st each switched proof-structure is a tree, i.e. a connected and acyclic graph.*

We recall:

**Theorem 6.** *A proof structure $\Theta$ is a Danos-Regnier proof net with conclusions $\Gamma$ iff there exists a proof $\pi$ of the sequent $\vdash \Gamma$ in multiplicative linear logic. Furthermore, one may define a correspondence between proofs and proof-nets.*

## 7.2 Sequent calculus style / Reduction by small steps

Let $\leadsto_C$ be the transitive closure of the following relation defined on literals of a formula $C$: let $u$ and $v$ be two literals of $C$, $u \leadsto_C v$ iff $\exists G, H$ st $C$ is of the form $C[G[u^\perp] \invamp H[v]]$. We note $u \leadsto v$ when there is no ambiguity.

**Definition 6 (Small step reduction rule).** *Let $\to$ be the reduction relation given by:*
$$C[\phi[x] \otimes \psi[x^\perp]] \to C[\phi[\psi[\perp]]]$$

*if $\forall v \in fv(\psi[.]), v \not\leadsto_C x$.*

**Theorem 7 (Stability).** *Let $x$ and $x^\perp$ appear only once in $C$, if $C[\phi[x] \otimes \psi[x^\perp]] \vdash$ is provable and the reduction rule applies, then $C[\phi[\psi[\perp]]] \vdash$ is provable.*

*Proof.* The proof consists in five main steps. In a first part we sketch the steps and give clues for the simplest ones. In a second part, we detail the more complex second step.

– The five steps:
  1. a proof $\pi$ of $C[\phi[x] \otimes \psi[x^\perp]] \vdash$ exists then $\pi$ is of the form:

     $$\vdots$$
     $$\mathcal{S}, \phi[x], \psi[x^\perp] \vdash$$
     $$\vdots \quad (1)$$
     $$C[\phi[x] \otimes \psi[x^\perp]] \vdash$$

     where $\pi'$ is the subproof of $\pi$ with conclusion $\mathcal{S}, \phi[x], \psi[x^\perp] \vdash$, where $\mathcal{S}$ is a multiset of formulae. The proof is done by recurrence (note that the language is only multiplicative).

  2. a proof $\pi'$ of $\mathcal{S}, \phi[x], \psi[x^\perp] \vdash$ exists then there exists $\pi''$, proof of $\mathcal{S}, \phi[x], \psi[x^\perp] \vdash$ of the form:

     $$\vdots$$
     $$\mathcal{T}, x, \psi[x^\perp] \vdash$$
     $$\vdots \quad (2)$$
     $$\mathcal{S}, \phi[x], \psi[x^\perp] \vdash$$

     we delay the proof of this step below.

  3. a proof $\rho$ of $\mathcal{T}, x, \psi[x^\perp] \vdash$ exists then a proof $\nu$ of $\mathcal{T}, \psi[\perp] \vdash$ exists: the variable $x$ appears only once in the proof $\rho$, then this proof is of the form (the language is only multiplicative):

     $$\cdots \quad \overline{x, x^\perp \vdash}$$
     $$\vdots$$
     $$\mathcal{T}, x, \psi[x^\perp] \vdash$$

     Hence, by recurrence one may build a proof $\nu$ of the form:

     $$\cdots \quad \overline{\perp \vdash}$$
     $$\vdots$$
     $$\mathcal{T}, \psi[\perp] \vdash$$

4. a proof $\nu$ of $\mathcal{T}, \psi[\bot] \vdash$ exists, then a proof of $\mathcal{S}, \phi[\psi[\bot]] \vdash$ exists: one completes the proof $\nu$ with the inference steps used in (2) (this may be proved by recurrence on the number of inference steps).
5. a proof of $\mathcal{S}, \phi[\psi[\bot]] \vdash$ exists then a proof of $C[\phi[\psi[\bot]]] \vdash$ exists: as the step before, inference steps in (1) may now be applied to get the result.

– proof of the second step: briefly speaking, we transform the proof $\pi'$ by commuting rules with subformulae of $\psi[]$ or $\mathcal{S}$ as principal formula with rules with $\phi[]$ as principal formula in order to decompose $\phi[]$ before $\psi[]$ (reading the proof bottom-up). The constraint makes it possible. The demonstration is done by induction on the height of the proof and in three steps:

1. We mark formulae in the proof $\pi'$ in the following way:
   - we mark $\bar{A}$ each formula $A$ that is a subformula of a formula containing a literal $w$ st $w \in \psi[]$ or $\exists v \in \psi[], v \rightsquigarrow w$.
   - we mark $\tilde{A}$ each formula $A$ that is a subformula of a formula containing a literal $w$ st $w \in \phi[]$ or $w \rightsquigarrow x$. Note that $\psi[]$ cannot be marked with $\tilde{\cdot}$ because of the constraint. Suppose that a formula $A$ has two marks then it must be of the form $\bar{A}_1 \otimes \tilde{A}_2$ and $A_1$ and $A_2$ have no other marks because of the constraint. Because $\otimes_L$ is asynchronous we can decompose first such a formula (that is different from $\psi[]$ as we noted above). Hence we consider for the following that a formula is not marked simultaneously $\tilde{\cdot}$ and $\bar{\cdot}$.
   - we mark $\dot{A}$ a formula unmarked by the two previous rules.

2. We prove now a one step inversion property (remember we "read" the proof bottom-up):
   - if $\pi'$ contains a rule on a formula marked $\bar{\cdot}$ followed by a rule $\otimes_L$ on a formula marked $\tilde{\cdot}$ or $\dot{\cdot}$, one may commute the two inferences: $\otimes_L$ is asynchronous and the marks being different, the formulae are distinct.
   - if $\pi'$ contains a rule $\mathclap{\ensuremath{\mathfrak{P}}}_L$ on a formula marked $\bar{\cdot}$ followed by a rule $\mathclap{\ensuremath{\mathfrak{P}}}_L$ on a formula marked $\tilde{\cdot}$ or $\dot{\cdot}$, one may commute the two inferences: use either the associativity of $\mathclap{\ensuremath{\mathfrak{P}}}_L$ if the inferences are done on the same formula or the fact that contexts are splitted independently.
   - if $\pi'$ contains a rule $\otimes_L$ on a formula marked $\bar{\cdot}$ followed by a rule $\mathclap{\ensuremath{\mathfrak{P}}}_L$ on a formula marked $\tilde{\cdot}$, one may commute the two inferences: there are really two cases,

$$\cfrac{\cfrac{\bar{a}, \bar{B}_1, \tilde{\alpha}, \tilde{C}_1, \dot{D}_1 \vdash \quad \bar{b}, \bar{B}_2, \tilde{\beta}, \tilde{C}_2, \dot{D}_2 \vdash}{\bar{a}, \bar{b}, \bar{B}_1, \bar{B}_2, \widetilde{\alpha \mathclap{\ensuremath{\mathfrak{P}}} \beta}, \tilde{C}_1, \tilde{C}_2, \dot{D}_1, \dot{D}_2 \vdash} \mathclap{\ensuremath{\mathfrak{P}}}}{a \otimes b, \bar{B}_1, \bar{B}_2, \widetilde{\alpha \mathclap{\ensuremath{\mathfrak{P}}} \beta}, \tilde{C}_1, \tilde{C}_2, \dot{D}_1, \dot{D}_2 \vdash} \otimes_L$$

or

$$\cfrac{\cfrac{\bar{B}_1, \tilde{\alpha}, \tilde{C}_1, \dot{D}_1 \vdash \quad \bar{a}, \bar{b}, \bar{B}_2, \tilde{\beta}, \tilde{C}_2, \dot{D}_2 \vdash}{\bar{a}, \bar{b}, \bar{B}_1, \bar{B}_2, \widetilde{\alpha \mathclap{\ensuremath{\mathfrak{P}}} \beta}, \tilde{C}_1, \tilde{C}_2, \dot{D}_1, \dot{D}_2 \vdash} \mathclap{\ensuremath{\mathfrak{P}}}}{a \otimes b, \bar{B}_1, \bar{B}_2, \widetilde{\alpha \mathclap{\ensuremath{\mathfrak{P}}} \beta}, \tilde{C}_1, \tilde{C}_2, \dot{D}_1, \dot{D}_2 \vdash} \otimes_L$$

But one of two premises does not contain formulae marked ?̃ (the one without $x$ and $x^\perp$), hence the first one is not allowed. Then the commutation property follows.
- other cases are treated easily.
3. we end with a classical induction on the height of the proof to show that decomposition of ?̃-formulae may be done before decomposition of ?̃-formulae. □

**Theorem 8 (Inverse Stability).** *If $C[\phi[\psi[\bot]]] \vdash$ is provable, then $C[\phi[x] \otimes \psi[x^\perp]] \vdash$ is provable.*

*Proof.* If $C[\phi[\psi[\bot]]] \vdash$ is provable then there exists a proof of the following form (where $\mathcal{S}$ and $\mathcal{T}$ are multisets of formulae):

$$\overline{\bot \vdash}$$
$$\vdots$$
$$\mathcal{T}, \psi[\bot]] \vdash$$
$$\vdots$$
$$\mathcal{S}, \phi[\psi[\bot]] \vdash$$
$$\vdots$$
$$C[\phi[\psi[\bot]]] \vdash$$

As $x, x^\perp \vdash$ is provable, one may prove by induction on the height of the proof (applying the previous inference steps) that there exists a proof of the following form:

$$\overline{x, x^\perp \vdash}$$
$$\vdots$$
$$\mathcal{T}, \psi[x^\perp], x \vdash$$
$$\vdots$$
$$\dfrac{\mathcal{S}, \psi[x^\perp], \phi[x] \vdash}{\mathcal{S}, \psi[x^\perp] \otimes \phi[x] \vdash}$$
$$\vdots$$
$$C[\phi[x] \otimes \psi[x^\perp]] \vdash$$

□

**Theorem 9 ((small steps) Correctness criterion).** *Let $M$ be a closed module, $M$ is c-correct iff $t(M) \to^* \bot$, where $\to^*$ is the transitive closure of $\to$, quotiented by the neutrality of $\bot$ wrt $\invamp$.*

*Proof.* The proof relies on stability and inverse stability. Let $M$ be a closed module.

- $\bot$ is the type of a c-correct module, hence by inverse stability, if $t(M)$ reduces to $\bot$, $M$ is c-correct.
- Suppose $M$ is c-correct. The proof is done by induction on the number of variables appearing in $M$.

- If $t(M)$ does not contain variables, $t(M)$ is of the form $\bot$, $\bot \parr F$ or $\bot \otimes F$ where $F$ is built with $\bot, \parr$ and $\otimes$. The last case is not provable hence contradicting the fact that $M$ is c-correct. The second case is equivalent to $F$ (neutrality of $\bot$ wrt $F$). We conclude by induction on the number of symbols appearing in $t(M)$.
- If $t(M)$ contains variables. Remember that $t(M)$ is of the form $\bigotimes (\parr_i h_i^\bot \parr \parr_k \bigotimes_{j_k} c_k^{j_k})$. As $t(M) \vdash$ is provable, there exists a polarized proof where each step is a complete decomposition of a formula corresponding to an elementary module (one of the elements of the tensor). Let us now consider one of the last (when we read the proof bottom-up) polarized decomposition, it has the following form:

$$\frac{\ldots \quad \overline{x_l, x_l^\bot \vdash} \quad \ldots}{\ldots, x_l, \ldots, \parr_l x_l^\bot \vdash}$$

Remark that the remainder of the proof does not contain with $\parr_l x_l^\bot$ as one premise an application of (i) a $\otimes$ except for the last step because $M$ is a composition of elementary modules, (ii) a $\parr$ because the proof is polarized. If $t(M) \vdash$ is obtained by application of a tensor rule, then the reduction rule applies considering whatever literal $x_l$. Otherwise, the previous proof follows with a polarized step $\otimes$ then $\parr$:

$$\frac{\dfrac{\ldots \quad \overline{x_l, x_l^\bot \vdash} \quad \ldots}{\dfrac{\ldots, x_l, \ldots, \parr_l x_l^\bot \vdash}{\ldots, y \parr \bigotimes_j x_l^j, \ldots, \parr_j x_l^{j\bot} \parr \parr_{j'} x_l^{j'\bot}, \cdots \vdash}}}{\vdots \quad t(M) \vdash} \quad \{j\} \cup \{j'\} = \{l\}$$

Then $t(M)$ is of the form $C[\psi[\bigotimes_j x_l^j] \otimes (\parr_j x_l^{j\bot} \parr \parr_{j'} x_l^{j'\bot})]$ for each $l$. The reduction criterion is satisfied for each $x_l^j$ hence the reduction rule applies. Finally, we note that the formula we get after reduction (and possibly using the neutrality of $\bot$ wrt $\parr$) is the type of the composition of elementary modules. □

### 7.3 (standard) Proof-net style / Small steps

The correctness of (closed) modules may be tested by means of a contraction criterion. In this subsection, we consider that modules are represented as proof structures and we use extensively the Danos-Regnier criterion to test that a proof-structure is correct. Let $\leadsto_\Theta$ be the transitive closure of the following relation defined on literals of a proof-structure $\Theta$: let $u$ and $v$ be two literals of $\Theta$, $u \leadsto_\Theta v$ iff $u^\bot$ and $v$ are in the same subtree with root $\otimes$ of the formula corresponding to $\Theta$. We note $u \leadsto v$ when there is no ambiguity. In the following, we consider proof-structures modulo neutrality of the constant $\mathbf{1}$ and associativity of connective $\parr$.

**Definition 7 (Small step reduction rule).** *Let $\to$ be the reduction relation given by:*

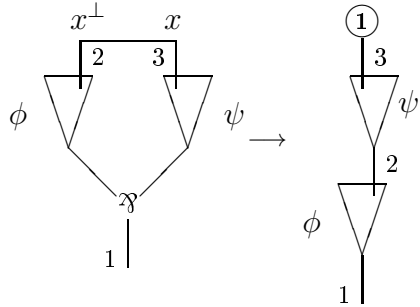

*when $\forall v$ a literal of $\psi$, $v \not\rightsquigarrow x^\perp$*

**Theorem 10 (Stability).** *Let $\Theta$ be a proof-structure, if $\Theta$ is DR-correct and $\Theta \to \Theta'$ then $\Theta'$ is DR-correct.*

*Proof.* Let $\Theta$ be a proof-net satisfying the criterion of the theorem, let $x$ be the variable involved in the reduction rule and $\phi$ and $\psi$ as given by the rule. We have to prove acyclicity and connexity of $\Theta'$. We prove connexity of $\Theta'$ by contradiction. Suppose there exists a switching function $s'()$ for $\Theta'$ and $u$ in $\phi$ and $v$ in $\psi$ are not connected. We consider the switching function $s()$ in $\Theta$ extending $s'()$ with a link for the connective $\Re$ (use $\Re_L$). As $\Theta$ is DR-correct, $u$ and $v$ are linked in $s(\Theta)$. $x$ must be in the path, otherwise the path remains in $s'(\Theta')$ and $u$ and $v$ are linked. So there is a path $u \ldots x^\perp x \ldots v$, hence there is a path in $s'(\Theta')$ from $u$ to point 2 (the previous path cannot go twice by $x$ otherwise there is a cycle). Finally, because we choose the link $\Re_L$, in $s(\Theta)$ either the path from $v$ to the root of $\psi$ $r_\psi$ does not pass through $x$ and there is a path in $s'(\Theta')$ from the point 2 to $v$ (and $u$ and $v$ are linked), or there exists $w$ in $\psi$ and a path $r_\psi \ldots w w^\perp \ldots u \ldots x^\perp$. In this latter case, $w \rightsquigarrow x^\perp$ and there is a contradiction.

The demonstration of acyclicity is done by induction on the height of $\psi$. We first prove commutation properties:

- suppose the principal connective of $\phi$ is a $\Re$, then $\Theta$ is of the left form below. $\Re$ being associative, $\Theta$ is correct iff the right proof-structure $\Pi$ below is correct.

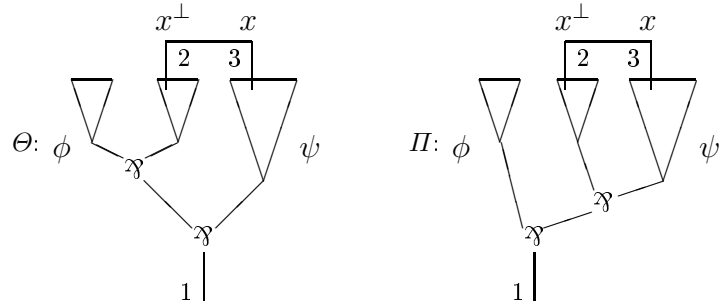

- suppose the principal connective of $\phi$ is a $\otimes$, then $\Theta$ is of the left following form. We prove the commutation (i.e. acyclicity of the right proof-structure $\Pi$) by contradiction.

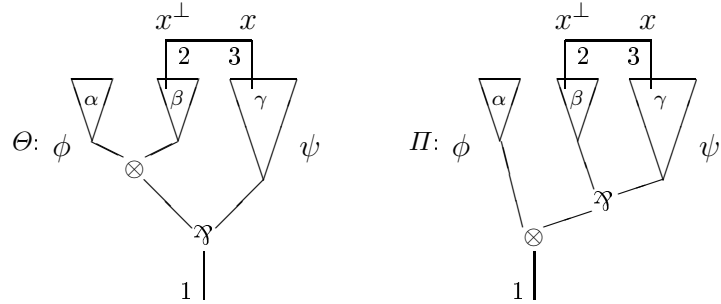

Let us suppose that there exists a cycle in $\Pi$ in one of the switched proof-structures, say $s(\Pi)$. Remark that the switched proof-structures issued from the sub-structures $\alpha$, $\beta$ and $\gamma$ of $\Pi$ are acyclic as this is true by hypothesis. We prove the result by studying cases wrt the switch chosen for $\invamp$:

- if there is a cycle, this one goes through at least one of the two nodes labelled $\otimes$ and $\invamp$. Otherwise, this cycle is also a cycle in $\Theta$.

- suppose $s(\invamp) = \invamp_R$:

  * if there exists a cycle $\pi$ going through the node labelled $\invamp$ but not through $\beta$, then $\exists v$ literal of $\gamma$ and $w$ literal of $\alpha$ st $\pi$ is of the form $w \ldots r_\alpha \otimes \invamp r_\gamma \ldots v \ldots w$. Hence, there exist switched proof-structures for $\Theta$ with path $r_\gamma \ldots v \ldots w \ldots r_\alpha$ that do not go through $\invamp$ and $\otimes$. Amongst these, one can choose at least one switched proof-structure with the path $v \ldots w \ldots r_\alpha \otimes r_\beta x^\perp$. Hence, $v$ is a literal of $\psi$ and $v \rightsquigarrow x^\perp$. Contradiction.

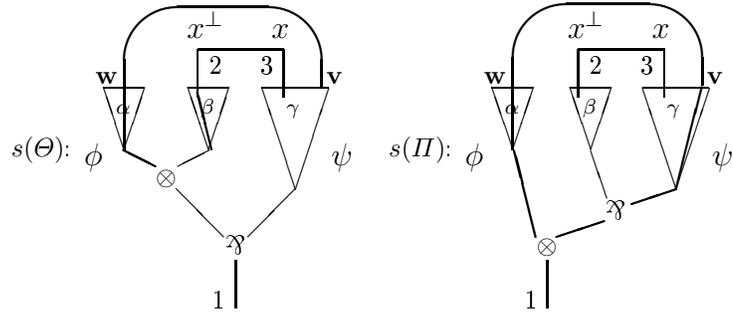

* if there exists a cycle $\pi$ going through the node labelled $\invamp$ and through $\beta$, then either this cycle goes through $x$ or not:

  · if $\pi$ goes through $x$ then there exists $v \in \beta, v \neq x^\perp$ and $s(\Pi)$ looks like the right figure. Hence one may choose a switch for $\Theta$ that defines a cycle as in the left one.

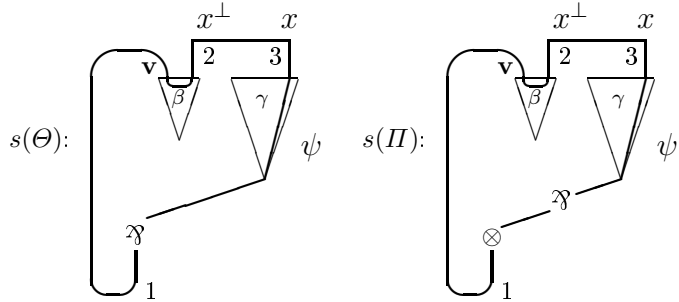

  · if $\pi$ does not go through $x$ then there exist $v$ and $w$ distinct from $x^\perp$ in $\beta$ and $u \neq x$ in $\gamma$ st the cycle looks like the figure in the right. We define a path linking $u$ to $x^\perp$ in a switched proof-structure of $\Theta$ depending on the main connective of the formula labelling the node joining $v$ and $x^\perp$. If this connective is $\otimes$ then one may define a switch joining $u$ and $x^\perp$ via $v$, hence $u \rightsquigarrow x^\perp$. If the connective is a $\invamp$, one may define a switch joining also $u$ and $x^\perp$ as in the figure on the left.

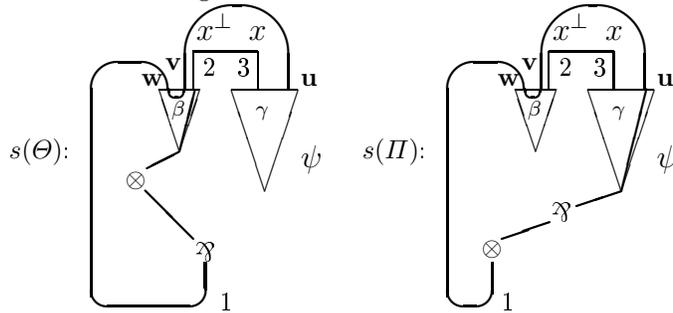

* if there exists a cycle $\pi$ not going through the node labelled $\invamp$, then this is also a cycle wrt a switched proof-structure of $\Theta$.

- suppose $s(\mathfrak{N}) = \mathfrak{N}_L$. From a switch for $\Pi$ including a cycle that goes through $\mathfrak{N}$, one may define a switch for $\Theta$ including a cycle that goes through $\otimes$, hence contradiction with acyclicity of $\Theta$.
- Finally, let us suppose that $\phi$ contains only one node labelled $x^\perp$, the following reduction is stable under correctness:

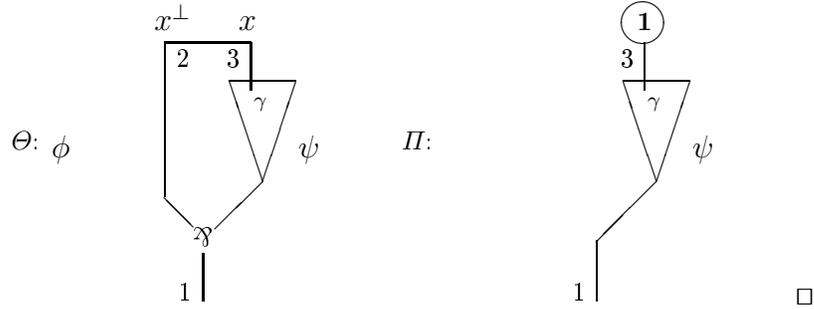

**Theorem 11 (Inverse Stability).** *Let $\Theta$ be a proof-structure, if $\Theta \to \Theta'$ and $\Theta'$ is DR-correct then $\Theta$ is DR-correct.*

*Proof.* Let $\Theta$ and $\Theta'$ be as in the theorem. Suppose there exists $s()$ and $s(\Theta)$ is not connected. Let $u$ and $v$ be variables not connected in $s(\Theta)$. We can suppose that $u \in \phi$ and $v \in \psi$: in other cases, either connexity remains as graph is unchanged or the cases reduce to the one we consider. Define $s'()$ as $s()$ without the switch for the $\mathfrak{N}$ (bottom of the figure for $\Theta$). $s'()$ is a switching function for $\Theta'$. As $\Theta'$ is DR-correct, $u$ and $v$ are connected in $s'(\Theta')$. Either the path from $v$ to $u$ goes through point 2 or not. If it does not go through point 2, it goes through point 1 and exits $\psi$ by a variable $w$. Then $w \in \psi$ and $w \rightsquigarrow x^\perp$ (principal connectives must be $\otimes$ otherwise the path in $s'(\Theta')$ is not valid). Contradiction. It it goes through point 2: there is a path in $s'(\Theta')$ from $u$ to point 2, then in $s(\Theta)$ there is a link from $u$ to $x^\perp$, there is a path in $s'(\Theta')$ from $v$ to **1** (point 3) because $\Theta'$ is DR-correct, then in $s(\Theta)$ there is a link from $v$ to $x$. Hence $u$ and $v$ are connected.

Suppose there exists $s()$ and there is a cycle in $s(\Theta)$. Let $s'()$ be $s()$ without the switch for the $\mathfrak{N}$ (bottom of the figure for $\Theta$). Either the cycle goes through $x$ or not. If it does not go through $x$, either the switch for $\mathfrak{N}$ is a $\mathfrak{N}_L$ or a $\mathfrak{N}_R$. If it is a $\mathfrak{N}_L$, then one has immediately a cycle in $s'(\Theta')$. If it is a $\mathfrak{N}_R$ and does not go through $\phi$, one has still a cycle in $s'(\Theta')$. Otherwise one can modify $s()$ st one keeps the cycle and there is a direct path from the root of $\phi$ to one of the variables of $\phi$ in the cycle. In this case there is a cycle choosing $\mathfrak{N}_L$ that does not go through $x$ and we are done. Finally, if it goes through $x$, suppose first that it does not go through $\mathfrak{N}$, there must exist $u \neq x^\perp$ exiting $\phi$ and $v \neq x$ exiting $\psi$ in the cycle. There is a path from $u$ to $v$ that does not go through $\phi$ and $\psi$, and a path from $u$ to point 2 that does not go through $x$. Hence it is possible to define a switching function st there is a cycle in $\Theta'$. If it goes through $\mathfrak{N}_R$ then there must exist a variable $u$ in $\phi$ where the cycle exits $\phi$ to go to point 1 without going through $\phi$, hence the cycle in $\Theta'$ is obvious. If it goes through $\mathfrak{N}_L$, it goes from point 2 to point 1 and from point 1 to $\psi$ entering at a variable, say $v$ without going through $\phi$. We can define a switching function for $\Theta'$ st there is a cycle. □

**Theorem 12 ((small steps) Correctness criterion).** *Let $M$ be a closed BM, $M$ is correct iff $M^* \to^* \mathbf{1}$.*

*Proof.*
– Obviously the proof-structure $\mathbf{1}$ is DR-correct. Furthermore there is no other DR-correct proof-structures without variables modulo neutrality of $\mathbf{1}$ wrt $\otimes$.
– Suppose $M^*$ is DR-correct, has still variables and is not reducible. For each variable $v$, the pattern of the left hand side of the rule is satisfied (otherwise it is easy to define a switch that gives a cycle), hence the condition on $\leadsto$ is not satisfied: there exists a variable $v'$ blocking $v$, i.e. $v' \leadsto v^\perp$. Let $v_0$ be a variable. From the previous remark and the fact that the number of variables is finite, one can define a circular list of variables $v_0, \ldots, v_n$ st $v_i$ blocks $v_{i+1}$ ($i$ modulo $n$). One may then define a cycle on the proof-structure that does not intersect with $\otimes$ as minimal connectives. Then we can define a switching function st there is a cycle in the switched structure. Contradiction.
– the rewriting rule is stable wrt DR-correctness and reducibility must occur until there is no variable then if $M$ is correct then $M^* \to^* \mathbf{1}$.
– the rewriting rule is stable wrt the inverse rewriting rule, then if $M^* \to^* \mathbf{1}$ then $M$ is correct. □

### 7.4 Open modules: properties of the contraction relation without completion

**Proposition 7 (stability).** *Let $M$ be an open module st $M \to_w M'$. The following properties are stable wrt contraction rules:*

1. *border set, i.e. $b(M') = b(M)$,*
2. *acyclicity,*
3. *connexity,*
4. *connectablility,*
5. *c-correction,*
6. *o-correction.*

*Proof.* By the trivial cases of 1., 2. and 3. we have the stability of c-correction. Regarding stability of connectability, remark that if $M$ is connectable by $\overline{M}$ then from 1. and 3. $M'$ is connectable by $\widetilde{M} \circ M'$. By the way o-correction is stable wrt contraction rules. □

**Proposition 8 (inverse stability).** *Let $M$ be an open module st $M \to_w M'$. The properties of the proposition 7 are stable wrt inverse contraction rules.*

The proof uses the same arguments.